\documentclass[aip,12pt,a4paper,preprint,amsmath,amssymb,groupedaddress]{revtex4-2} 

\usepackage{bm}
\usepackage{graphicx}

\usepackage{times}
\usepackage{mathptmx}

\DeclareSymbolFont{newfont}{OML}{cmm}{m}{it}
\DeclareMathSymbol{\Varrho}{3}{newfont}{37}

\newcommand{\p}{\partial}

\newcommand{\ave}[1]{{\left<#1\right>}}
\newcommand{\vek}[1]{{\mathbf#1}}

\newcommand{\rhos}{{\rho_\text{s}}}

\newcommand{\Cs}{{C_\text{s}}}
\newcommand{\bkappa}{{\pmb\kappa}}

\newcommand{\taud}{\ensuremath{\tau_\text{d}}}

\newcommand{\tauw}{\ensuremath{\tau_\text{w}}}

\newcommand{\aave}{\ensuremath{\ave{a}}}
\newcommand{\aoave}{\ensuremath{\ave{a_{0}}}}

\newcommand{\vave}{\ensuremath{\ave{v}}}

\newcommand{\ellave}{\ensuremath{\ave{\ell}}}

\newcommand{\Phiave}{\ensuremath{\ave{\Phi}}}
\newcommand{\Psiave}{\ensuremath{\ave{\Psi}}}

\newcommand{\Phirms}{\ensuremath{\Phi}_\text{rms}}
\newcommand{\Psirms}{\ensuremath{\Psi}_\text{rms}}

\newcommand{\amin}{\ensuremath{a_\text{min}}}
\newcommand{\amax}{\ensuremath{a_\text{max}}}
\newcommand{\aomin}{\ensuremath{a_{0\text{min}}}}

\newcommand{\vmin}{\ensuremath{v_\text{min}}}
\newcommand{\vmax}{\ensuremath{v_\text{max}}}
\newcommand{\ellmin}{\ensuremath{\ell_\text{min}}}
\newcommand{\ellmax}{\ensuremath{\ell_\text{max}}}

\newcommand{\taumin}{\ensuremath{\tau_\text{min}}}
\newcommand{\taumax}{\ensuremath{\tau_\text{max}}}

\newcommand{\taup}{\ensuremath{\tau_\shortparallel}}

\newcommand{\Eqref}[1]{Eq.~\eqref{#1}}
\newcommand{\Eqsref}[1]{Eqs.~\eqref{#1}}
\newcommand{\Figref}[1]{Fig.~\ref{#1}}

\newcommand{\Secref}[1]{Sec.~\ref{#1}}

\newcommand{\Appref}[1]{App.~\ref{#1}}

\begin{document}

\title{Stochastic modelling of blob-like plasma filaments in the scrape-off layer: Continuous velocity distributions}

\author{J.~M.~Losada}
\email{juan.m.losada@uit.no}

\author{O.~Paikina}
\email{olga.paikina@uit.no}

\author{O.~E.~Garcia}
\email{odd.erik.garcia@uit.no}

\affiliation{Department of Physics and Technology, UiT The Arctic University of Norway, N-9037 Troms{\o}, Norway}

\date{\today}

\begin{abstract}
A stochastic model for a superposition of uncorrelated pulses with a random distribution of amplitudes, sizes, and velocities is analyzed. The pulses are assumed to move radially with fixed shape and amplitudes decreasing exponentially in time due to linear damping. The pulse velocities are taken to be time-independent but randomly distributed. The implications of a broad distribution of pulse amplitudes and velocities, as well as correlations between these, are investigated. Fast and large-amplitude pulses lead to broad and flat average radial profiles with order unity relative fluctuations in the scrape-off layer. For theoretically predicted blob velocity scaling relations, the stochastic model reveals average radial profiles similar to the case of a degenerate distribution of pulse velocities but with more intermittent fluctuations. The average profile e-folding length is given by the product of the average pulse velocity and the linear damping time due to losses along magnetic field lines. The model describes numerous common features from experimental measurements and underlines the role of large-amplitude fluctuations for plasma-wall interactions in magnetically confined fusion plasmas.
\end{abstract}

\maketitle

\tableofcontents

\clearpage

\section{Introduction}

One of the main challenges to overcome to achieve economically efficient fusion energy by magnetic confinement is to control plasma and heat transport in the scrape-off layer (SOL) and plasma interactions with material surfaces.\cite{stangeby_plasma_2000,fundamenski_power_2014,krasheninnikov_edge_2020,militello_bonundary_2023} 
A common approach to describe the cross-field transport is based on mixing length estimates, assuming scale separation, small relative fluctuations and that the fluxes are proportional to local profile gradients.\cite{connor_invariance_1988,connor_comparison_1999,connor_relationship_2001,weiland_review_2016,stangeby_tutorial_2000,stangeby_modeling_2002,naulin_turbulent_2007,terry_scrape-off_2007}
On this basis, effective diffusivities, or a combination of effective diffusion and radial convection, have been used to characterize the cross-field transport in the SOL.\cite{stangeby_tutorial_2000,stangeby_modeling_2002,naulin_turbulent_2007,terry_scrape-off_2007,pigarov_tokamak_2002,umansky_numerical_2011}
Such an approach has been shown to be ill-founded by analysis of both experimental measurement data and first-principles-based turbulence simulations.\cite{labombard_particle_2001,garcia_turbulent_2007,garcia_fluctuations_2007,krasheninnikov_strongly_2009,zweben_edge_2015}
Indeed, turbulence studies indicate that the cross-field transport is dominated by the radial motion of field-aligned plasma filaments.\cite{krasheninnikov_recent_2008,garcia_blob_2009,dippolito_convective_2011} It has since become clear that the cross-field transport of particles and heat in the SOL depends on the amplitudes, sizes, velocities and frequency of occurrence of such blob-like structures, as well as correlations between these quantities. Therefore, a statistical approach is required to describe the resulting fluctuations and transport.\cite{garcia_stochastic_2012,kube_convergence_2015,theodorsen_level_2016,garcia_auto-correlation_2017,theodorsen_statistical_2017,theodorsen_level_2018,theodorsen_probability_2018,garcia_stochastic_2016,militello_scrape_2016,militello_relation_2016,walkden_interpretation_2017,militello_two-dimensional_2018,losada_stochastic_2023}

Theoretical investigations of isolated blob-like structures in the scrape-off layer have revealed the physical mechanism for their radial motion and the scaling of their velocity with amplitude, size and plasma parameters.\cite{garcia_mechanism_2005,madsen_influence_2011,wiesenberger_radial_2014,olsen_temperature_2016,pecseli_solvable_2016,held_influence_2016,wiesenberger_unified_2017,held_beyond_2023,garcia_radial_2006,kube_velocity_2011,manz_filament_2013,omotani_effects_2015,walkden_dynamics_2016,krasheninnikov_scrape_2001,dippolito_cross-field_2002,garcia_interchange_2006,myra_collisionality_2006,kube_velocity_2011,easy_three_2014,halpern_three-dimensional_2014,omotani_effects_2015,easy_investigation_2016,walkden_dynamics_2016} Scaling theory, as well as numerical simulations of seeded filament structures, have revealed that the radial blob velocity increases with the amplitude. In the so-called inertial regime, there is a linear dependence, while in the sheath-dissipative regime the velocity scales as the square root of the amplitude. In both regimes, there is a saturation of the amplitude dependence when the filament amplitude is much higher than the background level. Such a positive correlation between filament amplitudes and velocities has also been identified in experimental measurement data.\cite{theodorsen_scrape-off_2016}

The fluctuating plasma parameters in the SOL can be described as a superposition of uncorrelated pulses. Detailed analysis of experimental measurement data has demonstrated that these pulses have an exponential shape and that the pulse amplitudes are exponentially distributed.\cite{garcia_burst_2013,garcia_intermittent_2013,garcia_intermittent_2015,theodorsen_scrape-off_2016,kube_fluctuation_2016,garcia_sol_2017,garcia_intermittent_2018,kube_statistical_2019,kuang_plasma_2019,kube_comparison_2020,bencze_characterization_2019,zweben_temporal_2022,ahmed_strongly_2023} Such a statistical description, referred to as a filtered Poisson process, predicts the fluctuations to follow a Gamma distribution with the shape parameter given by the degree of pulse overlap, that is, the ratio of the average pulse duration and waiting times. This model has been found to be an excellent description of single-point measurements of intermittent fluctuations at the boundary of magnetically confined plasmas.\cite{garcia_burst_2013,garcia_intermittent_2013,garcia_intermittent_2015,theodorsen_scrape-off_2016,kube_fluctuation_2016,garcia_sol_2017,garcia_intermittent_2018,kube_statistical_2019,kuang_plasma_2019,kube_comparison_2020,bencze_characterization_2019,zweben_temporal_2022,ahmed_strongly_2023}

Recently, the filtered Poisson process was extended to describe the radial motion of blob-like plasma filaments and the resulting radial plasma profiles.\cite{garcia_stochastic_2016,militello_relation_2016,militello_scrape_2016,walkden_interpretation_2017,militello_two-dimensional_2018,losada_stochastic_2023} Analytical expressions have been derived for the cumulants and the lowest-order statistical moments of the process, elucidating how these depend on the distribution of pulse velocities and their correlations with the amplitudes.\cite{losada_stochastic_2023} When all pulses have the same velocity, the average radial profile decreases exponentially with radius with an e-folding length given by the product of the radial velocity and the parallel loss time. In Ref.~\onlinecite{losada_stochastic_2023}, closed-form expressions for the cumulants were obtained for the case of a discrete uniform distribution of pulse velocities independent of the amplitude distribution. In the present contribution, these results are extended to broad and continuous velocity distributions, and correlations between pulse velocities and amplitudes are investigated. Closed analytical expressions are in general not available, so the cumulants and the lowest-order moments are calculated numerically.

It is here demonstrated that fast and large-amplitude pulses may lead to nearly flat average radial profiles with order unity relative fluctuations in the scrape-off layer. For theoretically predicted blob velocity scaling relations, the stochastic model reveals average radial profiles similar to the case of a degenerate distribution of pulse velocities but with more frequent large-amplitude fluctuations. This holds for the velocity scaling relations in both the inertial and sheath-dissipative regimes. The average profile e-folding length is given by the product of the average pulse velocity and the linear damping time due to losses along magnetic field lines, while the relative fluctuation level and the skewness and flatness moments increase radially outwards. The predictions compare favourably with experimental measurements and underline the role of large-amplitude fluctuations for plasma-wall interactions in magnetically confined fusion plasmas.

This paper is organized as follows. In \Secref{sec:scaling} the theoretical velocity scaling for isolated blob-like filament structures is reviewed, with particular emphasis on the correlations between pulse amplitudes and velocities. The stochastic model describing a superposition of filaments is presented and discussed in \Secref{sec.model}. In \Secref{sec.vdist} we present the profiles obtained for various discrete and continuous velocity distributions. For all cases presented in this section, the amplitudes are taken to be exponentially distributed and independent of the velocities. In \Secref{sec.adist} a correlation between pulse amplitudes and velocities is investigated using a truncated exponential distribution of pulse amplitudes. A discussion of the results and the main conclusions are given in \Secref{sec.discussion}. The change of the pulse amplitude distribution with radial position is discussed in \Appref{app.axi}. Finally, the role of a distribution of pulse sizes is discussed in \Appref{app.Pell}.

\section{Blob velocity scalings}\label{sec:scaling}

The stochastic modelling to be presented in the following section includes correlations between the amplitude, size and velocity of blob-like plasma filaments. This is provided by basic mathematical descriptions of filament structures in magnetized plasmas which are reviewed here. Their evolution is determined by the plasma vorticity equation, which for a non-uniformly magnetized plasma is given by\cite{garcia_two-dimensional_2006,krasheninnikov_recent_2008}
\begin{equation}\label{eq.vorticity}
    \vek{b}\cdot\nabla\times\left( \rho\,\frac{\text{d}\vek{V}}{\text{d}t} \right) = B\vek{B}\cdot\nabla\left( \frac{J_\parallel}{B} \right) + 2\vek{b}\cdot\bkappa\times\nabla P ,
\end{equation}
where $\rho$ is the mass density, $\vek{V}$ is the fluid velocity, $\vek{b}=\vek{B}/B$ is the unit vector along the magnetic field $\vek{B}$, $J_\parallel$ is the $\vek{B}$-parallel electric current density, $\bkappa=(\vek{b}\cdot\nabla)\vek{b}$ is the magnetic curvature vector and $P$ is the plasma pressure. In the absence of parallel currents, an order of magnitude estimate with $\nabla_\perp\sim1/\ell$, $\text{d}/\text{d}t\sim V/\ell$, $\kappa\sim1/R$, where $R$ is the magnetic field radius of curvature, and $\nabla P/\rho\sim C_\text{s}^2\triangle n/\ell(N+\triangle n)$ immediately gives the inertial velocity scaling\cite{garcia_mechanism_2005,madsen_influence_2011,wiesenberger_radial_2014,olsen_temperature_2016,pecseli_solvable_2016,held_influence_2016,wiesenberger_unified_2017,held_beyond_2023,garcia_radial_2006,kube_velocity_2011,manz_filament_2013,omotani_effects_2015,walkden_dynamics_2016}
\begin{equation}\label{Vinertial}
    \frac{V}{\Cs} \sim \left( \frac{2\ell}{R}\frac{\triangle n}{N+\triangle n} \right)^{1/2}
\end{equation}
where $\Cs$ is the sound speed, $\ell$ is the cross-field blob size, and $\triangle n$ is the blob amplitude above the background particle density $N$. In this regime, the velocity increases with the square-root of the cross-field size. For small relative amplitudes, $\triangle n/N\ll1$, the velocity has a square-root dependence on the amplitude, $V\sim(\triangle n/N)^{1/2}$, while for large relative amplitudes, $\triangle n/N\gg1$, there is a saturation and the velocity becomes independent of the amplitude.

At the divertor targets, the boundary condition for the parallel electric current density is $J_\parallel=en\Cs[1-\exp(e\phi/T_\text{e})]$, where $\phi$ is the plasma electric potential relative to the sheath and $n$ is the particle density. With $V\sim\phi/B\ell$, balancing the parallel current and interchange terms on the right hand side of \Eqref{eq.vorticity} gives the sheath-dissipative velocity scaling\cite{krasheninnikov_scrape_2001,dippolito_cross-field_2002,garcia_radial_2006,myra_collisionality_2006,kube_velocity_2011,manz_filament_2013,easy_three_2014,halpern_three-dimensional_2014,omotani_effects_2015,easy_investigation_2016,walkden_dynamics_2016}
\begin{equation}\label{Vsheath}
    \frac{V}{\Cs} \sim \frac{2L_\parallel\rho_\text{s}^2}{R\ell^2} \frac{\triangle n}{N+\triangle n} ,
\end{equation}
where $\rhos$ is the sound gyro radius. This scaling holds for field-aligned filaments that are electrically connected to the target sheaths. In this regime, the velocity is inversely proportional to the blob size and scales linearly with the relative amplitude for small amplitudes.

More generally, in the intermediate regime between the inertial and sheath-dissipative scalings, an order of magnitude estimate gives a quadratic equation for the blob velocity,\cite{kube_velocity_2011} 
\begin{equation} \label{eq:order_of_mag}
    \frac{V^2}{\ell^2} - c_1 \frac{g}{\ell}\frac{\triangle n}{N+\triangle n} + c_2 \sigma_\phi V\ell = 0 ,
\end{equation}
where the exponential function in the sheath dissipation terms has been linearized. Here $c_1$ and $c_2$ are parameters that depend on the relative blob amplitude $\triangle n/N$. The positive root of \Eqref{eq:order_of_mag} is given by
\begin{equation}\label{eq.vscaling}
    \frac{V}{V_*} = \frac{c_2}{2}\left(\frac{\ell}{\ell_*}\right)^3 \left[ -1 + \left( 1 + \frac{c_1}{c_2}\frac{4\ell_*^5}{\ell^5}\frac{\triangle n}{N+\triangle n} \right)^{1/2} \right] ,
\end{equation}
where we have defined a characteristic blob size
\begin{equation}
    \frac{\ell_*}{\rhos} = \left( \frac{2L_\parallel^2}{R\rhos} \right)^{1/5} ,
\end{equation}
and a corresponding characteristic blob velocity
\begin{equation}
    \frac{V_*}{\Cs} = \left( \frac{8\rho_\text{s}^2L_\parallel}{R^3} \right)^{1/5} .
\end{equation}
In Ref.~\onlinecite{kube_velocity_2011}, \Eqref{eq.vscaling} was found to be an excellent parametrization of the maximum velocity in numerical simulations of isolated blob structures. For large blob sizes, $\ell/\ell_*\gg1$, the inertial scaling from \Eqref{Vinertial} is recovered,
\begin{equation}
    \frac{V}{V_*} \sim \left( c_1\,\frac{\ell}{\ell_*}\frac{\triangle n}{N+\triangle n} \right)^{1/2} .
\end{equation}
Due to the absence of sheath currents, which leads to dissipation of large length scales, the velocity increases with the blob size $\ell$. For small blob sizes, $\ell/\ell_*\ll1$, the sheath dissipative scaling from \Eqref{Vsheath} follows,
\begin{equation}
    \frac{V}{V_*} \sim \frac{c_1}{c_2}\left( \frac{\ell_*}{\ell} \right)^2 \frac{\triangle n}{N+\triangle n} .
\end{equation}
The dominating sheath currents lead to strong dissipation of large length scales, resulting in a velocity that decreases with increasing size. These blob velocity scaling regimes appear to accurately describe experimental measurements in magnetized plasmas, based on both electric probe measurements and gas puff imaging diagnostics, as well as numerical simulations of SOL filaments and turbulence.\cite{theiler_blob_2011,kube_blob_2013,fuchert_influence_2013,fuchert_blob_2014,zweben_blob_2016,carralero_recent_2017,decristoforo_interactions_2020,han_estimating_2023}

\section{Stochastic model}\label{sec.model}

Consider a stochastic process with a random variable $\Phi_K$  given by a super-position of $K$ uncorrelated pulses,
\begin{equation} \label{eq.PhiK}
    \Phi_K(x,t) = \sum_{k=1}^{K(T)} \phi_k(x,t-s_k) .
\end{equation}
where each pulse $\phi$ has an arrival time $s$ at the reference position $x=0$ and satisfies the evolution equation
\begin{equation}\label{eq.phik}
    \frac{\p\phi}{\p t} + v\,\frac{\p\phi}{\p x} + \frac{\phi}{\taup} = 0 .
\end{equation}
Here and in the following, we suppress notation of the $k$-index except when summation over pulses is explicit. In \Eqref{eq.phik}, $v$ is the pulse velocity and $\taup$ is a constant linear damping time describing drainage due to particle motion along magnetic field lines in the scrape-off layer. All pulses are assumed to have the same shape with the initial condition given by
\begin{equation}
    \phi(x,0) = a_{0}\varphi\left(\frac{x}{\ell}\right) ,
\end{equation}
where $\ell$ is the pulse size and $a_{0}$ is the pulse amplitude at the reference position $x=0$. The general solution of \Eqref{eq.phik} can be written as
\begin{equation}
    \phi(x,t) = A(t)\varphi\left(\frac{x-vt}{\ell}\right) ,
\end{equation}
where the pulse amplitude decreases exponentially in time due to the linear damping,
\begin{equation}
    A(t) = a_{0}\exp\left( -\frac{t}{\taup} \right) .
\end{equation}
The stochastic process can accordingly be written as
\begin{equation}
    \Phi_K(x,t) = \sum_{k=1}^{K(T)} a_{0k}\exp\left( -\frac{t-s_k}{\taup} \right)\varphi\left(\frac{x-v_k (t-s_k)}{\ell_k}\right) .
\end{equation}
The random variables in this model are as follows:
\begin{itemize}
    \item[$K$:] The total number of pulses at the reference position $x=0$ during a time interval of duration $T$, taken to be Poisson distributed with average waiting time $\tauw$,
        \begin{equation}\label{eq.poisson}
        P_K(K;T) = \frac{1}{K!}\left(\frac{T}{\tauw}\right)^K\exp\left(-\frac{T}{\tauw} \right) .
        \end{equation}
    \item[$s$:] The pulse arrival time at the reference position $x=0$, which is uniformly distributed on the time interval $[-T/2,T/2]$ and thus given by
\begin{equation}
        P_s(s) = \frac{1}{T} .
    \end{equation}
    \item[$a_0$:] The pulse amplitude at the reference position $x=0$, with marginal probability density function $P_{a_0}(a_0)$.
    \item [$v$:] The pulse velocity, assumed to be positive definite and time-independent with marginal probability density function $P_v(v)$.
    \item [$\ell$:] The pulse size, taken to be the same for all pulses unless otherwise stated (specifically in \Appref{app.Pell}).
\end{itemize}
A distribution of pulse sizes does not influence the radial variation of the cumulants as long as they are not correlated with the pulse amplitudes or velocities.\cite{losada_stochastic_2023}

\begin{figure*}[tb]
\centering
\includegraphics[width=0.5\textwidth]{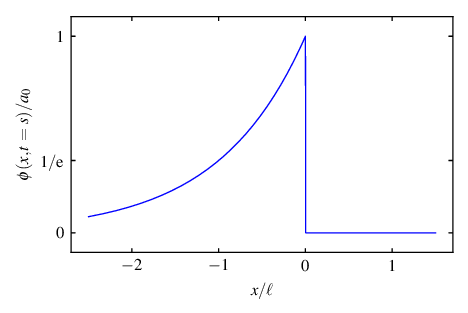}
\caption{Radial variation of a one-sided exponential pulse at the arrival time $s$ with amplitude $a$ and size $\ell$. The pulse moves along the $x$-axis with velocity $v$.}
\label{fig.pulseshape}
\end{figure*}

In the following, we will consider a one-sided exponential pulse function,
\begin{equation}\label{eq.phiexp}
    \varphi(\theta) = \Theta(-\theta)\exp(\theta) ,
\end{equation}
where $\Theta$ denotes the unit step function,
\begin{equation}
    \Theta(\theta) = \begin{cases} 1 , & \theta\geq0 \\ 0 , & \theta<0 . \end{cases}
\end{equation}
The solution of \Eqref{eq.phik} for an individual pulse can then be written as
\begin{equation}
    \phi(x,t) = a_{0} \exp\left( -\frac{t}{\taup} \right)\exp\left(\frac{x-vt}{\ell}\right)\Theta\left(-\frac{x-vt}{\ell}\right) .
\end{equation}
The one-sided pulse function is presented in \Figref{fig.pulseshape}, describing the radial motion of a blob-like structure with a steep front and a trailing wake.

A pulse $\phi$ will arrive at position $\xi$ at time $s_{\xi}=s+\xi/v$. The superposition of pulses at this position for a general pulse function $\varphi(\theta)$ can thus be written as
\begin{equation}\label{eq.Phi_K_xi}
    \Phi_K(\xi,t) = \sum_{k=1}^{K(T)} a_{\xi k} \exp\left( - \frac{t-s_{\xi k}}{\taup} \right) \varphi\left( -\frac{v_k(t-s_{\xi k})}{\ell_k} \right) ,
\end{equation}
where the pulse amplitudes are given by
\begin{equation}\label{eq.axi}
    a_{\xi} = a_{0} \exp\left( - \frac{\xi}{v\taup} \right) .
\end{equation}
When all pulses have the same velocity, the amplitudes will have the same probability density function at position $\xi$ but with a mean value that decreases exponentially with radius. For the one-sided exponential pulse function defined by \Eqref{eq.phiexp}, this takes a particularly simple form,
\begin{equation}\label{eq.Phi_K_xi2}
    \Phi_K(\xi,t) = \sum_{k=1}^{K(T)} a_{\xi k} \varphi\left( \frac{t-s_{\xi k}}{\tau_k} \right) ,
\end{equation}
where the pulse duration $\tau$ is given by the harmonic mean of the linear damping and radial transit times,
\begin{equation}\label{eq.tauk.onesided}
    \tau = \frac{\taup\ell}{\taup v + \ell} .
\end{equation}
The average pulse duration $\taud=\ave{\tau}$ is obtained by averaging over the distribution of pulse sizes and velocities. Equation~\eqref{eq.Phi_K_xi} demonstrates that the process can be regarded as a superposition of pulses with amplitudes that decrease radially outwards due to linear damping.

The lowest-order statistical moments of the process can be derived from the cumulants, which are the coefficients in the expansion of the logarithm of the characteristic function.
In the case of time-independent velocities and an exponential pulse function, the cumulants of the process $\Phi_K$ become\cite{losada_stochastic_2023}
\begin{equation}\label{eq.cumulants}
    \kappa_n(x) = \frac{1}{n\tauw}\,\ave{ a_0^n\tau \exp{\left( -\frac{nx}{v\taup} \right)} } .
\end{equation}
The averages in this equation are to be performed over the random variables $a_0$, $v$ and $\ell$, which in general are described by a joint probability distribution. When closed-form expressions for the cumulants are not available, it is straightforward to perform the integrations in \Eqref{eq.cumulants} numerically. However, the presence of slow pulses may lead to issues with the existence of cumulants and moments of the process for negative $x$. With the amplitudes specified as $a_0$ at the reference position $x=0$, slow pulses with very long radial transit times will have excessively large upstream amplitudes. This results in divergence of cumulants if the velocity probability distribution does not decrease sufficiently fast for small velocities. For this reason, lower truncated amplitude and velocity distributions will be considered in the following sections. Further discussions on the existence of cumulants are given in Ref.~\onlinecite{losada_stochastic_2023}.

From the cumulants, we can derive expressions for the lowest-order moments. Specifically, the mean value is given by the first order cumulant, $\Phiave=\kappa_1(x)$, which has e-folding length $L_\Phi(x)=1/(\text{d}\ln{\Phiave}/\text{d}x)$. The variance is given by the second order cumulant, $\Phirms^2=\langle{(\Phi-\Phiave)^2}\rangle=\kappa_2(x)$. We further define the skewness and flatness moments respectively by
\begin{equation}
    S_\Phi(x) = \frac{\kappa_3(x)}{\kappa_2^{3/2}(x)} ,
    \qquad
    F_\Phi(x) = \frac{\kappa_4(x)}{\kappa_2^2(x)} .
\end{equation}
For a normally distributed random variable, these two moments vanish. In the following section, we will use the relative fluctuation level $\Phirms/\Phiave$ and the skewness and flatness moments to quantify the presence of large-amplitude fluctuations in the process.

\begin{figure}[tb]
\centering
\includegraphics[width=0.5\textwidth]{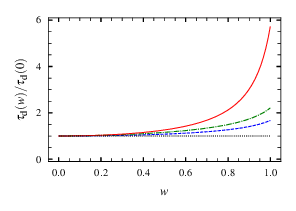}
\caption{Average pulse duration as a function of the width parameter $w$ for discrete uniform (red full line), continuous uniform (blue dashed line) and truncated exponential (green dashed-dotted line) distribution of pulse velocities. For all cases, the normalized linear damping time is $\ave{v}\taup/\ell=10$.}
\label{fig.taud}
\end{figure}

A variation in the pulse velocities also implies a distribution of pulse durations as described by \Eqref{eq.tauk.onesided}. When all pulses have the same size, the pulse duration distribution is related to the velocity distribution by the standard rules for the transformation of random variables,
\begin{equation}\label{eq.ptau}
    P_\tau(\tau) = \frac{\ell}{\tau^2}\,P_v\left( \frac{\ell}{\tau}-\frac{\ell}{\taup} \right) .
\end{equation}
Since the pulse duration is inversely proportional to the velocity, the average duration is dominated by the small velocities in the case of a random velocity distribution. This is demonstrated by the results presented in \Figref{fig.taud}, which shows the average pulse duration as function of the width parameter of the velocity distribution when this is discrete uniform, continuous uniform and truncated exponential (the corresponding probability density functions are defined in the following section). The $1/2$ probability for low pulse velocities in the case of a discrete uniform velocity distribution results in a sixfold increase in the average duration as the width parameter approaches $1$. For the widest continuous velocity distributions, the average pulse duration increases by a factor of approximately two. It is larger for the truncated exponential velocity distribution since this has higher probability for low pulse velocities. The increase in the average pulse duration with the width parameter of the velocity distribution also increases the mean value of the process at the reference position, which for uncorrelated pulse amplitudes and durations is given by $\Phiave(0)=(\taud/\tauw)\aoave$. The relative fluctuation level as well as the skewness and flatness moments will be correspondingly reduced at $x=0$, as will be seen in the following sections.

\begin{figure}[tb]
\includegraphics[width=8cm]{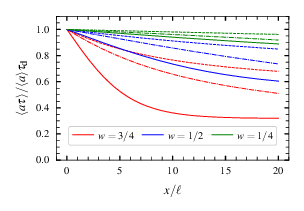}
\caption{Radial variation of the linear correlation between pulse amplitudes and durations for various width parameters $w$ for a discrete uniform (full line), continuous uniform (dashed line) and truncated exponential (dashed-dotted line) velocity distribution. For all cases, the amplitude distribution at the reference position $x=0$ is exponential and the normalized linear damping time is $\vave\taup/\ell=10$.}
\label{fig.teff-uncorr}
\end{figure}

When there is a distribution of pulse velocities, the amplitudes and durations will be correlated downstream even if they are independent at the reference position $x=0$. This is due to their dependence on pulse velocity, given by \Eqsref{eq.axi} and \eqref{eq.tauk.onesided}. Fast pulses have short radial transit times and less amplitude reduction due to linear damping. Therefore, $a_\xi$ increases with increasing velocity, while $\tau$ decreases with increasing velocity. 
In order to quantify this, we define an effective pulse duration which is weighted by the pulse amplitude and therefore varies with radial position, $\langle{a_\xi\tau}\rangle/\langle{a_\xi}\rangle$.
Where the average is to be taken over the distribution of pulse amplitudes, sizes and velocities. Similarly, we can define the normalized effective duration by dividing by the average duration, $\langle{a_\xi\tau}\rangle/\langle{a_\xi}\rangle\taud$. This is also a measure of the linear correlation between the pulse amplitude and duration. In \Figref{fig.teff-uncorr}, the radial variation of the normalized effective duration is presented  for various values of the width parameter of the velocity distribution when this is discrete uniform, continuous uniform and truncated exponential. This decreases radially outwards due to the anti-correlation between pulse amplitudes and durations. Accordingly, also the cumulants will decrease with increasing radial coordinate.

In the following two sections, it will be demonstrated how a distribution of pulse amplitudes and velocities and correlations between these influenc the lowest-order moments of the process. It should be noted that the cumulants can be written as $\kappa_n=\langle{a_\xi^n\tau}\rangle/n\tauw$. Thus, the correlation between pulse amplitudes and durations determines the radial variation of the cumulants and therefore also the moments of the process.

\section{Independent amplitudes and velocities}\label{sec.vdist}

In this section, we will investigate the case where pulse velocities and amplitudes are independent but both have a random distribution. The pulse amplitudes $a_0$ at the reference position $x=0$ are here assumed to have an exponential distribution, which for $a_0>0$ is given by
\begin{equation}\label{eq.pa.exp}
    \langle{a_0}\rangle P_{a_0}(a_0) = \exp{\left( - \frac{a_0}{\langle{a_0}\rangle} \right)} ,
\end{equation}
where $\langle{a_0}\rangle$ is the average pulse amplitude. For this distribution, the raw amplitude moments are given by $\langle{a_0^n}\rangle=n!\langle{a_0}\rangle^n$. In the following, various pulse velocity distributions will be considered. Pulse sizes are assumed to have a degenerate distribution, so all pulses have the same size $\ell$.

\subsection{Degenerate distribution}\label{sec:Pvdelta}

In the case of a degenerate distribution, all pulses have the same velocity $\vave$,
\begin{equation}
    P_v(v) = \delta(v-\ave{v}) .
\end{equation}
The cumulants given by \Eqref{eq.cumulants} then simplify to
\begin{equation}\label{eq.kappan.degenerate}
    \kappa_n(x) = \gamma_* (n-1)! \ave{a_0}^n \exp{\left( - \frac{nx}{\ave{v}\taup} \right) } ,
\end{equation}
where we have defined the intermittency parameter
\begin{equation}
    \gamma_* = \frac{\tau_*}{\tauw} 
\end{equation}
with the pulse duration
\begin{equation}
    \tau_* = \frac{\taup\ell}{\ave{v}\taup+\ell} .
\end{equation}
The cumulants given by \Eqref{eq.kappan.degenerate} describe a Gamma distribution with shape parameter $\gamma_*$ and scale parameter given by the radial pulse amplitude profile
\begin{equation}
    \aave(x)=\langle{a_0}\rangle\exp{\left( - \frac{x}{\ave{v}\taup} \right)} .
\end{equation}
The probability density function for positive $\Phi$ is the Gamma distribution
\begin{equation}
    \langle{a}\rangle P_{\Phi}(\Phi;x) = \frac{1}{\Gamma(\gamma_*)} \left(\frac{\Phi}{\langle{a}\rangle}\right)^{\gamma_*-1}\exp{\left(-\frac{\Phi}{\langle{a}\rangle}\right)} ,
\end{equation}
where $\Gamma$ denotes the Gamma function. The average radial profile is then given by\cite{garcia_stochastic_2016}
\begin{equation}
    \ave{\Phi}(x) = \gamma_* \langle{a_0}\rangle \exp{\left( - \frac{x}{\ave{v}\taup} \right)} ,
\end{equation}
which has e-folding length $L_\Phi=\vave\taup$. Moreover, the prefactor for the profile at the reference position is $\Phiave(0)=\gamma_*\aoave$. The lowest-order normalized moments are in this case radially constant,
\begin{equation}
\frac{\Phi_\text{rms}}{\ave{\Phi}} = \frac{1}{\gamma_*^{1/2}} ,
\quad
S_\Phi = \frac{2}{\gamma_*^{1/2}} ,
\quad
F_\Phi = \frac{6}{\gamma_*} .
\end{equation}
The mean value $\Phiave$ decreases exponentially with radius with an e-folding length given by the product of the radial pulse velocity $\ave{v}$ and the linear damping time $\taup$. The prefactor $\gamma_*\langle{a_0}\rangle$ is given by the product of the average pulse amplitude at the reference position $x=0$ and the ratio of the average pulse duration and waiting times. This defines the standard case in order to compare with a broad distribution of pulse velocities. A higher degree of pulse overlap, or larger $\gamma_*$, increases the mean value of the process and decreases the relative fluctuation level and intermittency of the process.

\subsection{Discrete uniform distribution}

The simplest non-degenerate case to consider is a discrete uniform distribution of pulse velocities, for which they can take two possible values with equal probability,\cite{losada_stochastic_2023}
\begin{equation}\label{eq.prob2v}
    P_v(v;w) = \frac{1}{2}\left[ \delta(v-\vmin) + \delta(v-\vmax) \right] ,
\end{equation}
where $\vmin=(1-w)\ave{v}$, $\vmax=(1+w)\ave{v}$, $\vave$ is the average pulse velocity and $w$ is the width parameter for the distribution with values in the range $0<w<1$. The limit $w\rightarrow0$ corresponds to the case of a degenerate distribution of pulse velocities discussed above.

The statistical properties of the process for this distribution were analyzed in detail in Ref.~\onlinecite{losada_stochastic_2023} and are summarized here in order to compare with continuous velocity distributions in the following subsections. The discrete uniform velocity distribution translates into a discrete uniform distribution of the pulse durations,
\begin{equation}\label{eq.Ptau.2v}
    P_\tau(\tau) = \frac{1}{2}\left[ \delta(\tau-\tau(\vmin)) + \delta(\tau-\tau(\vmax)) \right] ,
\end{equation}
with the velocity dependent pulse duration $\tau(v)$ given by \Eqref{eq.tauk.onesided}. The average pulse duration is given by integration over the discrete distribution,
\begin{equation}\label{eq.taud.2v}
    \taud = \frac{1}{2}\left[ \tau(\vmin)+\tau(\vmax)\right] .
\end{equation}
As shown in \Figref{fig.taud}, the average pulse duration increases with $w$ since this implies pulses with lower velocities and correspondingly longer radial transit times $\ell/\vmin$. In the limit $w\rightarrow1$, this results in nearly stagnant pulses and in the absence of linear damping results in a divergence of the average pulse duration and therefore the mean value of the process.\cite{losada_stochastic_2023}

The probability density function for the pulse amplitudes $a$ at position $x$ with the appropriate normalization is for $a>0$ given by
\begin{equation}\label{eq.Pa(x)}
    P_a(a;x) = \frac{1}{2\amin}\exp{\left(-\frac{a}{\amin}\right)} + \frac{1}{2\amax}\exp{\left(-\frac{a}{\amax}\right)} ,
\end{equation}
where we have defined the radial amplitude profile for the slow and fast pulses respectively by
\begin{align}
\amin(x) & = \langle{a_0}\rangle\exp{\left( -\frac{x}{\vmin\taup} \right)} ,
\\
\amax(x) & = \langle{a_0}\rangle\exp{\left( -\frac{x}{\vmax\taup} \right)} .
\end{align}
The radial profile of the average amplitude given by \Eqref{eq.axi} is a sum of two exponential functions,
\begin{equation}\label{eq.aave.2v}
    \aave(x) = \frac{\langle{a_0}\rangle}{2} \left[ \exp{\left( - \frac{x}{\vmin\taup} \right)} + \exp{\left( - \frac{x}{\vmax\taup} \right)} \right] .
\end{equation}
More generally, the cumulants for the process are given by
\begin{multline} \label{eq.cumulant.2v}
    \kappa_n(x) = \frac{\langle{a_0^n}\rangle}{2n\tauw}\,\\
    \left[ \tau(\vmin)\exp{\left( -\frac{nx}{\vmin\taup} \right)} + \tau(\vmax)\exp{\left( -\frac{nx}{\vmax\taup}\right)} \right] .
\end{multline}
In this case, the process $\Phi_K(x,t)$ can be considered as the sum of two filtered Poisson processes, each Gamma distributed with shape parameters $\taumin/2\tauw$ and $\taumax/2\tauw$, and scale parameters $\amin(x)$ and $\amax(x)$, corresponding respectively to the slow and fast pulses. Accordingly, the probability density function for the summed process is given by the convolution of the Gamma distributions for the two sub-processes. At the reference position $x=0$, the scale parameters for the two sub-processes are the same and the process is Gamma distributed with shape parameter $\taud/\tauw$ and scale parameter $\langle{a_0}\rangle$. At large radial positions the process is dominated entirely by the fast pulses and the probability density function is a Gamma distribution with shape parameter $\tau(\vmax)/2\tauw$ and scale parameter $\langle{a_0}\rangle\exp(-x/\vmax\taup)$.

\begin{figure*}[tb]
\centering
\includegraphics[width=\textwidth]{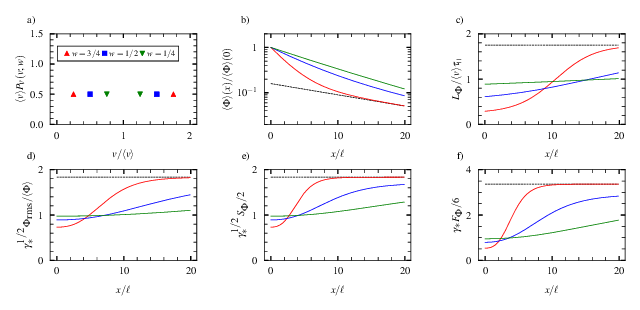}
\caption{Results for a discrete uniform distribution of pulse velocities and exponentially distributed pulse amplitudes at $x=0$. Plot panels show a) velocity distribution, b) average value, c) profile scale length, d) relative fluctuation level, e) skewness moment, and f) flatness moment for various values of the width parameter $w$ for the velocity distribution. All radial profiles are normalized with their values at the reference position $x=0$ for the case of a degenerate distribution of pulse velocities. For all radial profiles, the normalized linear damping time is $\ave{v}\taup/\ell=10$. The dashed lines represent the profiles for the process with only the fast pulses in the case $w=3/4$.}
\label{fig.full_plot_2v}
\end{figure*}

The discrete uniform velocity distribution is presented in \Figref{fig.full_plot_2v} together with the radial profile of the mean value, its e-folding length, the relative fluctuation level, and the skewness and flatness moments for various values of the width parameter $w$. All of these profiles are normalized by their values at the reference position $x=0$ for the standard case of a degenerate distribution of pulse velocities, discussed in \Secref{sec:Pvdelta}. The profiles therefore show how these moments are modified by a distribution of pulse velocities for fixed $\vave$ and $\gamma_*$. In the case of a wide separation of the pulse velocities, the radial profile of the mean value $\Phiave$ is steep at the reference position $x=0$ and becomes significantly flatter far downstream compared to the standard case. The relative fluctuation level and the skewness and flatness moments saturate at the level determined by the fast pulses only, indicated by the dashed lines in \Figref{fig.full_plot_2v}.

\subsection{Continuous uniform distribution}

Consider next a continuous uniform distribution of pulse velocities, which is non-zero only for velocities in the range $\vmin\leq v\leq\vmax$ and then given by
\begin{equation}\label{eq.pv.unif}
    \ave{v}P_{v}(v;w) = \frac{1}{2w} ,
\end{equation}
where $\vmin=(1-w)\vave$, $\vmax=(1+w)\vave$ and $w$ is the width parameter for the distribution with values in the range $0<w<1$. From \Eqsref{eq.ptau} and \eqref{eq.pv.unif} follows the probability distribution for the pulse duration $\tau$, which is finite in the interval $\taumin\leq\tau\leq\taumax$ and given by
\begin{equation}\label{eq.ptau.unif}
    P_{\tau}(\tau;w) = \frac{\ell}{2\ave{v}w \tau^2} ,
\end{equation}
where $\taumin=\taup\ell/(\vmax\taup+\ell)$ and $\taumax=\taup\ell/(\vmin\taup+\ell)$. The average pulse duration is then given by
\begin{equation}
    \taud = \frac{\ell}{2\vave w} \ln\left( \frac{\vave\taup(1+w)+\ell}{\vave\taup(1-w)+\ell} \right) .
\end{equation}
The variation of the average pulse duration with the width parameter is presented in \Figref{fig.taud}. As for the discrete case, the average duration increases with the width parameter $w$ due to the presence of slow pulses with long radial transit times. The power law scaling of the pulse durations in \Eqref{eq.ptau.unif} significantly influences temporal correlations and will result in long-range dependence for a wide distribution.\cite{garcia_auto-correlation_2017,korzeniowska_apparent_2023}

\begin{figure*}[tb]
\centering
\includegraphics[width=\textwidth]{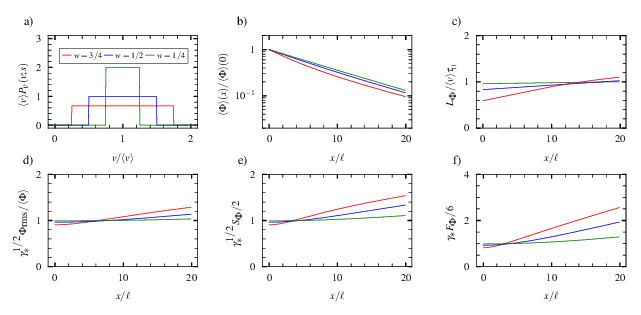}
\caption{Results for a continuous uniform distribution of pulse velocities and exponentially distributed pulse amplitudes at $x=0$. Plot panels show a) velocity distribution, b) average value, c) profile scale length, d) relative fluctuation level, e) skewness moment, and f) flatness moment for various values of the width parameter $w$ of the velocity distribution. All radial profiles are normalized with their values at the reference position $x=0$ for the case of a degenerate distribution of pulse velocities. For all radial profiles, the normalized linear damping time is $\ave{v}\taup/\ell=10$.}
\label{fig.full_plot_unif}
\end{figure*}

The continuous uniform velocity distribution as well as the radial profile of the lowest order moments of the process are presented in \Figref{fig.full_plot_unif}. The profiles have the same trend with increasing width parameter $w$ as for the case of a discrete uniform distribution of pulse velocities but the radial variation is weaker since there is a continuous range of allowed velocities. In particular, the mean value decreases nearly exponentially with radius for large radial positions, close to the case of a degenerate distribution of pulse velocities. However, there is a gradual increase in the relative fluctuation level and in particular the skewness and flatness moments with radius. This again demonstrates that the fluctuations become more intermittent for a broad distribution of pulse velocities.

\subsection{Truncated exponential distribution}\label{sec.vdist.texp}

\begin{figure*}[tb]
\centering
\includegraphics[width=\textwidth]{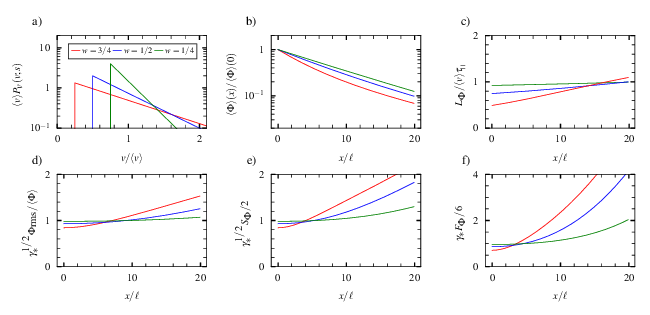}
\caption{Results for a lower truncated exponential distribution of pulse velocities and exponentially distributed pulse amplitudes at $x=0$. Plot panels show a) velocity distribution, and radial profiles of b) average value, c) profile scale length, d) relative fluctuation level, e) skewness moment, and f) flatness moment for various values of the width parameter $w$ of the velocity distribution. All radial profiles are normalized with their values at the reference position $x=0$ in the case of a degenerate distribution of pulse velocities. For all radial profiles, the normalized linear damping time is $\ave{v}\taup/\ell=10$.}
\label{fig.prof_exp}
\end{figure*}

Finally, we will consider a lower truncated exponential distribution of pulse velocities, which is non-zero for velocities $v\geq\vmin=(1-w)\vave$ and then given by
\begin{equation}
    \vave P_{v}(v;w) = \frac{1}{w}\,\exp{\left( - \frac{v-(1-w)\vave}{w\vave} \right)} .
\end{equation}
The truncation parameter $w$ is effectively a width parameter, where the limit $w\rightarrow0$ corresponds to the case of a degenerate distribution and $w\rightarrow1$ corresponds to the standard exponential distribution. The truncated exponential distribution is presented in \Figref{fig.prof_exp}a) for various values of the width parameter $w$.

The pulse duration distribution is non-zero for $\tau\leq\taumax=\taup\ell/(\vmin\taup+\ell)$ and is then given by
\begin{equation}
    P_\tau(\tau;w) = \frac{\ell}{\vave \tau^2w}\,\exp\left( -\frac{\ell}{\vave\tau w} \right)\exp\left( \frac{1-w}{w}+\frac{\ell}{\vave\taup w} \right) .
\end{equation}
The average pulse duration is
\begin{equation}
    \taud = \frac{\ell}{\vave w}\,\exp\left( \frac{(1-w)\vave\taup+\ell}{\vave\taup w} \right) \text{E}_1\left( \frac{(1-w)\vave\taup+\ell}{\vave\taup w} \right) ,
\end{equation}
where $\text{E}_1$ denotes the exponential integral function. This is presented as a function of the width parameter $w$ in \Figref{fig.taud}, showing an increase with a factor slightly larger than two in the limit $w\rightarrow1$ for $\vave\taup/\ell=10$. This is due to the dominant contribution of slow pulses for a wide distribution of pulse velocities. The mean value of the process will be correspondingly increased at the reference position. However, for large radial positions, the fast pulses dominate the process and the effective duration is substantially reduced, as shown in \Figref{fig.teff-uncorr}.

The radial profiles of the lowest order moments are presented in \Figref{fig.prof_exp} for various values of the width parameter for the truncated exponential velocity distribution. 
These radial profiles have the same trend as for the case of a continuous uniform distribution of pulse velocities but the radial variation of the mean value deviates more from exponential for a wide distribution due to the abundance of slow pulses. However, the relative fluctuation level and the skewness and flatness moments are much higher at large radial positions than for a discrete uniform velocity distribution, demonstrating strong intermittency of the fluctuations due to fast pulses with short radial transit times.

\section{Correlated amplitudes and velocities}\label{sec.adist}

In this section, we consider the effect of a correlation between pulse velocities and amplitudes at the reference position $x=0$, in particular cases where these have a linear or square-root relationship, as suggested by the blob velocity scaling estimates presented in \Secref{sec:scaling}. Since vanishing pulse velocities are not allowed in the model, we first describe the effect of truncating the exponential amplitude distribution given by \Eqref{eq.pa.exp}.

\subsection{Truncated exponential distribution}

\begin{figure*}[tb]
\centering
\includegraphics[width=0.5\textwidth]{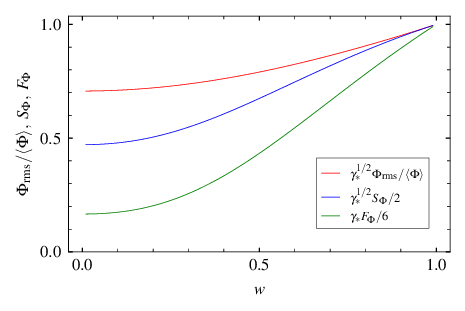}
\caption{Relative fluctuation level and skewness and flatness moments of the process for a degenerate distribution of pulse velocities and a lower truncated exponential amplitude distribution with minimum amplitude given by $a_\text{min}=(1-w)\langle{a_0}\rangle$ as a function of the width parameter $w$.}
\label{fig.prof_aexp}
\end{figure*}

Consider the process with a degenerate velocity distribution, $P_v(v)=\delta(v-\ave{v})$, and a lower truncated exponential distribution of pulse amplitudes, which is nonzero for $a_0\geq\aomin=(1-w)\aoave$ and given by
\begin{equation}
    \aoave P_{a_0}(a_0;w) = \frac{1}{w} \exp{\left( - \frac{a_0 - (1-w)\aoave}{w\aoave} \right)} .
\end{equation}
The cumulants can be calculated analytically and are given by
\begin{equation}
    \kappa_n(x) = \frac{\gamma}{n} \exp \left( -\frac{nx}{v\taup} \right) \sum_{i=0}^{n} \frac{n!}{(n-i)!} \left[ (1-w)\aave \right]^i \left[ w\ave{a} \right]^{n-i} .
\end{equation}
In the limit $w\rightarrow1$ this corresponds to a standard exponential amplitude distribution, which is the case described in \Secref{sec:Pvdelta}. In this limit, the raw amplitude moments are $\langle{a_0^n}\rangle=n!\aoave^n$ and the probability density function for the process is a Gamma distribution. In the limit $w\rightarrow0$, corresponding to a degenerate distribution of pulse amplitudes, the raw amplitude moments are given by $\langle{a_0^n}\rangle=\aoave^n$.
Accordingly, compared to the standard case with an exponential amplitude distribution, for the maximally truncated exponential amplitude distribution, the relative fluctuation level is reduced by a factor $1/\sqrt{2}$, the skewness moment by $\sqrt{2}/3$ and the flatness moment by $1/6$. The variation of these moments with the width parameter is presented in \Figref{fig.prof_aexp}. Each moment is normalized by its value in the case of a degenerate distribution of pulse velocities.

The mean value $\Phiave$ of the process does not vary with the width parameter $w$ as it only depends on the average amplitude $\langle{a_0}\rangle$ and $\gamma_*$. Note that the net effect of increasing the amplitude width parameter is to decrease intermittency, as is shown by a decrease in all higher-order moments. This is simply because there is less randomness in the process. In particular, when all pulses have the same amplitude and velocity, only the pulse arrival times and the number of pulses are randomly distributed.

\begin{figure*}[tb]
\centering
\includegraphics[width=\textwidth]{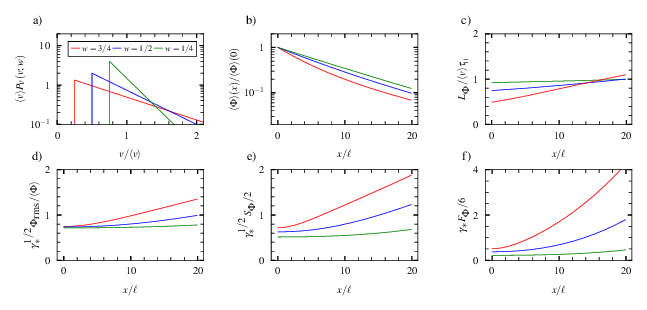}
\caption{Results for independent pulse amplitudes and velocities at $x=0$, both having a lower truncated exponential distribution with the same width parameter $w$. Plot panels show a) velocity distribution, and radial profiles of b) average value, c) inverse profile scale length, d) relative fluctuation level, e) skewness moment, and f) flatness moment for various values of the width parameter $w$. All radial profiles are normalized with their values at the reference position $x=0$ in the case of a degenerate distribution of pulse velocities. For all radial profiles, the normalized linear damping time is $\ave{v}\taup/\ell=10$.}
\label{fig.prof_exp_va}
\end{figure*}

It is furthermore of interest to consider the case in which pulse amplitudes and velocities are independent at the reference position $x=0$ but both have a lower truncated exponential distribution with the same width parameter $w$. This is in contrast to the case in which pulse amplitudes are lower truncated exponentially distributed and the velocities are given by a power law dependence on the pulse amplitudes, which will be analyzed in the following subsection. The resulting profiles are presented in \Figref{fig.prof_exp_va} for various values of the width parameter. These only differ from those resulting from a lower truncated exponential velocity distribution, shown in \Figref{fig.prof_exp}, by a constant factor. Thus, similar to the case discussed in \Secref{sec.vdist.texp}, the radial increase of the relative fluctuation level and the skewness and flatness moments are much more pronounced for a wide velocity distribution.

\subsection{Power law dependence}\label{sec.corr}

\begin{figure*}[tb]
\centering
\includegraphics[width=\textwidth]{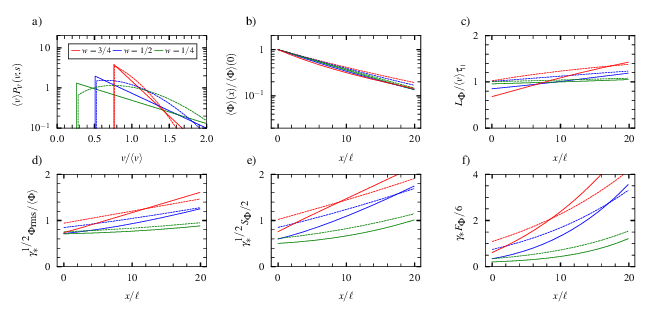}
\caption{Results for correlated pulse amplitudes and velocities, $v\sim a_0^\alpha$, with $\alpha=1/2$ (broken lines) and $\alpha=1$ (full lines), for a lower truncated exponential amplitude distribution with width parameter $w$ at $x=0$. Plot panels show a) marginal velocity distribution and radial profiles of b) average value, c) inverse profile scale length, d) relative fluctuation level, e) skewness moment, and f) flatness moment for various values of the width parameter $w$. All radial profiles are normalized with their values at the reference position $x=0$ in the case of a degenerate distribution of pulse velocities. For all radial profiles, the normalized linear damping time is $\ave{v}\taup/\ell=10$.}
\label{fig.prof_correxp}
\end{figure*}

As discussed in \Secref{sec:scaling}, blob velocity scaling theories demonstrate that the blob velocity depends on the amplitude and size. This motivates the study of cases in which the velocity is given by a power law dependence on the amplitude,
\begin{equation}
    \frac{v}{\ave{v}} = c_v\left( \frac{a_0}{\langle{a_0}\rangle} \right)^{\alpha} ,
\end{equation}
where the proportionality factor $c_v=\langle{a_0}\rangle^\alpha/\langle{a_0^\alpha}\rangle$ depends on the amplitude distribution.
The marginal velocity probability density function and the radial profile of the lowest order moments are presented in \Figref{fig.prof_correxp} for $\alpha=1/2$ and $\alpha=1$, and for various values of the width parameter for a lower truncated amplitude distribution. For all cases, the radial variation of the mean $\Phiave$ is nearly exponential with an e-folding length close to $\vave\taup$, the same as for the standard case with a degenerate velocity distribution. However, the higher order moments increase with radius and are much larger than in the standard case for a wide amplitude distribution. The intermittency is highest for the case $\alpha=1$ since this results in faster pulses with shorter radial transit times.

\begin{figure*}[tb]
\centering
\includegraphics[width=\textwidth]{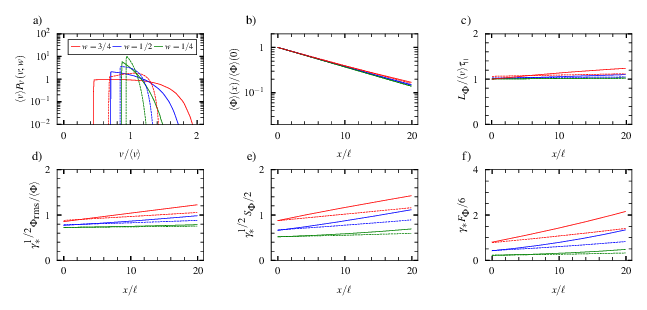}
\caption{Results for correlated pulse amplitudes and velocities, $v\sim[a_0/(1+a_0)]^\alpha$, with $\alpha=1/2$ (broken lines) and $\alpha=1$ (full lines), for a lower truncated exponential amplitude distribution with width parameter $w$ at $x=0$. Plot panels show a) marginal velocity distribution and radial profiles of b) average value, c) inverse profile scale length, d) relative fluctuation level, e) skewness moment, and f) flatness moment for various values of the width parameter $w$. All radial profiles are normalized with their values at the reference position $x=0$ in the case of a degenerate distribution of pulse velocities. For all radial profiles, the normalized linear damping time is $\ave{v}\taup/\ell=10$.}
\label{fig.prof_uncorr_va}
\end{figure*}

The blob velocity scaling theory suggests a saturation of the velocity dependence on pulse amplitudes when these become large compared to any background level, as described by \Eqref{eq.vscaling}. In order to investigate this, we finally consider the scaling relationship
\begin{equation}\label{eq.va-saturation}
    \frac{v}{\ave{v}} = c_v\left( \frac{a_0}{1+a_0} \right)^{\alpha} ,
\end{equation}
and a truncated exponential distribution for the pulse amplitudes $a_0$. The marginal velocity probability density function and the radial profile of the lowest order moments are presented in \Figref{fig.prof_uncorr_va} for $\alpha=1/2$ and $\alpha=1$, and for various values of the width parameter. Again we observe that for all cases, the radial variation of the mean $\Phiave$ is nearly exponential and similar to the standard case where all pulses have the same velocity. Thus, the e-folding length is determined by the average pulse velocity and has a weak dependence on the width of the distribution. The relative fluctuation level as well as the skewness and flatness moments increase radially outwards but not as strongly as for the power law scaling $v\sim a_0^\alpha$ due to the saturation for large pulse amplitudes described by \Eqref{eq.va-saturation}.

\section{Discussion and conclusions}\label{sec.discussion}

The stochastic modelling presented here shows that an exponential average profile follows from a superposition of uncorrelated pulses which all have the same radial velocity. In this standard case, the profile e-folding length is given by the product of the radial pulse velocity and the linear damping time due to particle motion along magnetic field lines in the SOL. At the reference position $x=0$, the prefactor for the profile is given by the product of the average pulse amplitude and the ratio of the average pulse duration and waiting times. The process is strongly intermittent with large-amplitude fluctuations when the latter ratio is small, implying little overlap of pulses. When the amplitudes are exponentially distributed, the process has a Gamma probability density with the ratio of the average pulse duration and waiting times as the shape parameter and the mean amplitude as the scale parameter. From this, it follows that the relative fluctuation level and the skewness and flatness moments are radially constant.\cite{losada_stochastic_2023}

A random distribution of pulse velocities significantly modifies the process. Firstly, it leads to longer average pulse duration times, thereby increasing the degree of pulse overlap. The relative fluctuation level, skewness and flatness are reduced close to the reference position due to longer pulse durations, as seen in 
Figs.~\ref{fig.full_plot_2v}--\ref{fig.prof_exp} and \ref{fig.prof_exp_va}--\ref{fig.prof_uncorr_va}.  However, there will also be an anti-correlation between pulse amplitudes and durations downstream from the reference position. This results in a shorter effective pulse duration with the process becoming dominated by the fast pulses which have short radial transit times and undergo less linear damping. As a result, the relative fluctuation level and the skewness and flatness moments increase radially outwards. In the simplest case of a discrete uniform velocity distribution, the average profile has a bi-exponential shape resembling the near and far SOL typically measured at the boundary of magnetically confined plasmas. 

For pulse velocities with continuous distributions but independent of pulse amplitudes, the average profile is in general non-exponential and has lower values than the standard case with a degenerate velocity distribution. However, the higher order moments, like the relative fluctuation level as well as the skewness and flatness moments, are generally higher and increase radially outwards. This is strongly amplified when there is a correlation between pulse amplitudes and velocities. The blob velocity scaling theory presented in \Secref{sec:scaling} suggests that the pulse velocity depends linearly on the amplitude in the inertial regime or as the square root of the amplitude in the sheath dissipative regime. In both cases, the average radial profile is close to that of the standard case where all pulses have the same velocity. The high average plasma density, long profile scale length and high relative fluctuation level in the SOL underline the importance of blob-like plasma filaments for plasma-wall interactions in magnetically confined fusion plasmas. 

The stochastic model presented here provides a unique statistical framework for analyzing and interpreting fluctuation measurements from the boundary region of magnetically confined plasmas as well as the output data from first principles-based turbulence simulations.\cite{garcia_stochastic_2012,kube_convergence_2015,theodorsen_level_2016,garcia_auto-correlation_2017,theodorsen_statistical_2017,theodorsen_level_2018,theodorsen_probability_2018,garcia_stochastic_2016,militello_scrape_2016,militello_relation_2016,walkden_interpretation_2017,militello_two-dimensional_2018,losada_stochastic_2023} It should be noted that there is no background or equilibrium plasma in the SOL according to the model. The mean value is entirely due to blob-like filament structures moving radially outwards. Numerous investigations have demonstrated both the underlying assumption of the model as well as its predictions are in excellent agreement with experimental measurements from single-point recording.\cite{garcia_burst_2013,garcia_intermittent_2013,garcia_intermittent_2015,theodorsen_scrape-off_2016,kube_fluctuation_2016,garcia_sol_2017,garcia_intermittent_2018,kube_statistical_2019,kuang_plasma_2019,kube_comparison_2020,bencze_characterization_2019,zweben_temporal_2022,ahmed_strongly_2023} This includes pulses arriving according to a Poisson process, an exponential pulse shape and an exponential distribution of pulse amplitudes. This results in a Gamma probability density function for the plasma fluctuations and a Lorentzian frequency power spectral density. In future work, the model will used to describe both fluctuation statistics, including pulse amplitude and velocity estimation, and the average radial profiles of the lowest order moments.

\appendix

\section{Amplitude distribution}\label{app.axi}

As discussed at the end of \Secref{sec.model}, the dominant mechanism for radial variation of the relative fluctuation level and the skewness and flatness moments is the correlation between pulse amplitudes and durations when there is a broad distribution of pulse velocities. However, the change in the amplitude distribution with radial position will also influence the moments of the process.

The pulse amplitude at any position $\xi$ is given by \Eqref{eq.axi}.
In the case of a random distribution of pulse velocities $v$, the amplitudes $a_{\xi}$ at position $\xi$ are given by the product of two random variables. The probability density of the product of two independent, non-negative random variables is given by the Mellin convolution of the two corresponding densities. Assuming that there is no correlation between amplitudes and velocities, the distribution of $a_{\xi}$ is thus given by
\begin{equation}
    P_{a_\xi} (a_\xi) = \int_{0}^{\infty} \frac{\text{d}a_0}{a_0} \, P_{a_0} (a_0) P_u \left(\frac{a_\xi}{a_0}\right)
\end{equation}
where $P_u$ is the probability distribution of the random variable $u=\exp{(-\xi/v\taup)}$.

As an example, consider an exponential distribution for the pulse amplitudes $a_0$ given by \Eqref{eq.pa.exp} and a truncated exponential distribution of pulse velocities. The exponential transform of the scaled reciprocal random variable $-\xi/v\taup$ results in a new random variable $u$, which has a probability density
\begin{equation}
     P_u (u) = \frac{\xi}{\taup w\ave{v}} \frac{1}{u \ln^2{u}} \exp{\left(\frac{\frac{\xi}{\taup\ln{u}}+(1-w)\ave{v}}{w\ave{v}}\right)},
\end{equation}
for $u$ in the range
\begin{equation}
    u_{\text{min}}=\exp{\left(-\frac{\xi}{\taup(1-w)\ave{v}}\right)}  \leq u<1.
\end{equation}
As the distributions $P_{a_0}$ and $P_{u}$ are known, we can determine $P_{a_\xi}$ at the radial position $\xi$. Note that the lower and upper limit of integration of $a_0$ is determined from the relation
\begin{equation}
    u=\frac{a_\xi}{a_0}, \quad a_0\geq 0, \quad u_{\text{min}}\leq u <1 ,
\end{equation}
and zero otherwise. Since $u<1$, the lower limit on $a_0$ is $a_\xi$. Also, since $u \geq u_{\text{min}}$, the upper limit on $a_0$ is $a_\xi/u_{\text{min}}$.

\begin{figure*}[ht]
\centering
\includegraphics[width=0.5\textwidth]{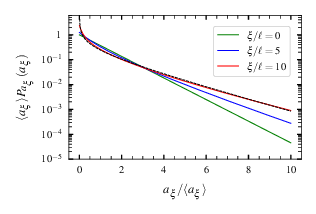}
\caption{Probability density function of the pulse amplitudes for a truncated exponential distribution of pulse velocities with width parameter $w=3/4$ at various radial positions in the case $\ave{v}\taup/\ell = 10$. The dashed line shows a Gamma distribution with shape parameter $\alpha=1/2$.}
\label{fig.ave_taud_correxp}
\end{figure*}

The amplitude probability distribution function $P_{a_{\xi}}$ is presented in \Figref{fig.ave_taud_correxp} for various radial positions $\xi$ for the truncated exponential velocity distribution with width parameter $w=3/4$. At $\xi=0$, the distribution is by definition exponential. The downstream amplitude distribution is peaked for small amplitudes since slow pulses have long radial transit times and are strongly depleted by linear damping. Far downstream, the distribution is well described by a Gamma distribution with a shape parameter $\alpha\leq1$,
\begin{equation}\label{eq.PaxiGamma}
    \langle{a_{\xi}}\rangle P_{a_{\xi}} = \frac{1}{\Gamma(\alpha)}\left( \frac{a_\xi}{\langle{a_\xi}\rangle} \right)^{\alpha-1} \exp{\left( -\frac{a_\xi}{\langle{a_\xi}\rangle} \right)} ,
\end{equation}
where the average amplitude $\langle{a_\xi}\rangle$ at the position $\xi$ is the scale parameter of the distribution. For this probability density the amplitude moments are given by $\langle{a_\xi^n}\rangle=\langle{a_\xi}\rangle^\alpha\Gamma(n+\alpha)/\Gamma(\alpha)$.

In order to quantify how the strongly peaked amplitude distribution modifies the process, consider a super-position of uncorrelated, exponential pulses\cite{garcia_auto-correlation_2017}
\begin{equation}
    \Psi_K(t) = \sum_{k=1}^{K(T)} a_{\xi k}\psi\left(\frac{t-s_k}{\taud}\right) ,
\end{equation}
with Gamma distributed amplitudes given by \Eqref{eq.PaxiGamma} and an exponential pulse function $\psi$ with fixed duration $\taud$. The lowest order moments for this process are readily calculated,\cite{theodorsen_probability_2018}
\begin{align}
    \Psiave & = \gamma\alpha\langle{a_\xi}\rangle , 
    \\
    \Psirms^2 & = \frac{1}{2}\,\gamma\alpha(1+\alpha)\langle{a_\xi^2}\rangle ,
    \\
    S_\Psi & = \frac{2^{3/2}}{3\gamma^{1/2}} \frac{\alpha+2}{[\alpha(1+\alpha)]^{1/2}} ,
    \\
    F_\Psi & = \frac{1}{\gamma}\frac{(2+\alpha)(3+\alpha)}{\alpha(1+\alpha)} .
\end{align}
When the shape parameter $\alpha$ is small, the relative fluctuation level and the skewness and flatness moments are very high due to the occasional appearance of large-amplitude pulses. In fact, in the limit $\alpha\rightarrow0$, the shape parameter has the same influence on these moments as the intermittency parameter $\gamma$ that determines the degree of pulse overlap. However, as discussed in \Secref{sec.model}, the main mechanism for the radial increase in relative fluctuation level and the skewness and flatness moments is the correlation between pulse amplitudes and durations. In Ref.~\onlinecite{losada_stochastic_2023}, this was explicitly demonstrated for a discrete uniform velocity distribution.

\section{Size distribution}\label{app.Pell}

\begin{figure*}[tb]
\centering
\includegraphics[width=\textwidth]{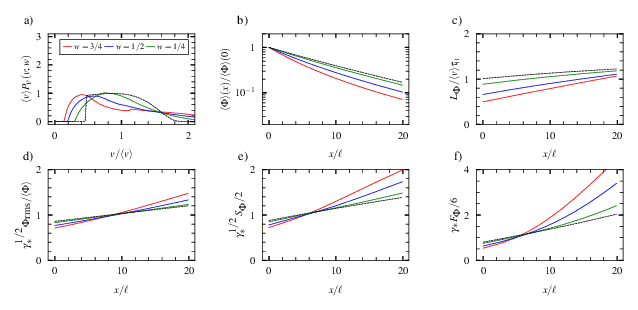}
\caption{Results for correlated pulse amplitudes, sizes and velocities given by \Eqref{eq.vscaling} with a truncated exponential amplitude distribution with width parameter $3/4$ and a uniform size distribution with width parameter $w$. Plot panels show a) marginal velocity distribution and radial profiles of b) average value, c) inverse profile scale length, d) relative fluctuation level, e) skewness moment, and f) flatness moment for various values of the width parameter $w$. All radial profiles are normalized with their values at the reference position $x=0$ in the case of a degenerate distribution of pulse velocities. The dashed lines show the profiles for a degenerate distribution of pulse sizes. For all radial profiles, the normalized linear damping time is $\ave{v}\taup/\ell=10$.}
\label{fig.prof_sheath_connected}
\end{figure*}

A random distribution of pulse sizes with modify the pulse durations and lead to more randomness in the process. Consider truncated exponentially distributed amplitudes with width parameter $3/4$, continuous uniformly distributed sizes $\ell$ in the range $\ellmin\leq\ell\leq\ellmax$,
\begin{equation}\label{eq.pl.unif}
    \ave{\ell}P_{\ell}(\ell;w) = \frac{1}{2w} ,
\end{equation}
where $\ellmin=(1-w)\ellave$ and $\ellmax=(1+w)\ellave$, and pulse velocities given by \Eqref{eq.vscaling}. The velocity distribution and resulting profiles of the lowest order moments are presented in \Figref{fig.prof_sheath_connected}.
Compared to \Figref{fig.prof_uncorr_va}, the velocity distribution is broader for a wide amplitude distribution. As a result, the e-folding length for the profile is shorter and the mean value lower than for the standard case with a degenerate velocity distribution. However the high velocity pulses result in larger fluctuation amplitudes and higher skewness and flatness moments for a wide size distribution.



\section*{Acknowledgements}

This work was supported by the UiT Aurora Centre Program, UiT The Arctic University of Norway (2020). The authors acknowledge discussions with A.~Theodorsen.

\bibliographystyle{apsrev4-1}
\bibliography{SOL}

\end{document}